\documentclass[%
 reprint,
nofootinbib,
 amsmath,amssymb,
 aps,
]{revtex4-2}

\usepackage{graphicx}
\usepackage{dcolumn}
\usepackage{bm}
\usepackage{aas_macros}
\usepackage[normalem]{ulem}
\usepackage{hyperref}
\usepackage{comment}

\usepackage{xcolor}

\newcommand{\as}[1]{#1}
\newcommand{\pa}[1]{#1}
\newcommand{\pas}[1]{#1}
 
\begin{document}


\title{Generalized Redundant Calibration of Radio Interferometers}

\author{Prakruth Adari}
\author{An\v{z}e Slosar}
\affiliation{Brookhaven National Laboratory, Upton NY 11973}
\affiliation{Physics and Astronomy Department, Stony Brook University, Stony Brook, NY 11794}

\begin{abstract}
Redundant calibration is a technique in radio astronomy that allows calibration of radio arrays whose antennas lie on a lattice by exploiting the fact that redundant baselines should see the same sky signal. Because the number of measured visibilities scales quadratically with the number of antennas but the number of unknowns describing the individual antenna responses and the available information about the sky scales only linearly with the array size, the problem is always over-constrained as long as the array is big and dense enough.  This is true even for non-lattice array configurations. In this work we study a generalized algorithm in which a per-antenna gain is replaced with a number of gains. We show that it can successfully fit data from an approximately redundant array on square lattice with pointing and geometry errors, but that the models parameters are difficult to link to the quantities of interest. We discuss the parameterization, limitations, and possible extensions of this algorithm.
\end{abstract}

\maketitle


\section{Introduction}

The 21\,cm emission from neutral hydrogen is promising to transform our understanding of the universe across the ages: from low-redshift observations of neutral hydrogen in galaxies through the epoch of reionization all the way to the dark ages at redshift $z\sim 100$ in the future.

The field received a major boost when it was realized that developments in computing and RF technology allow telescopes to be build almost entirely in software. A number of experiments were born, some already operating, such as CHIME\cite{CHIME}, Tianlai\cite{wu2016tianlai}, HERA\cite{2020AAS...23630804J}, MWA\cite{2020MNRAS.493.4711T}, \pa{PAPER\cite{Parsons_2010}, GMRT\cite{ryan2009global}}, some are under construction, including HIRAX\cite{HIRAX}, BINGO\cite{wuensche2019bingo},\pa{SKA-low\cite{dewdney2013ska1}}, and CHORD\cite{Vanderlinde:2019tjt} and some proposals for future very large facilities, such as PUMA \cite{2002.05072}.

Large interferometric radio arrays require a large number of calibration parameters. It was soon noticed that arrays of indistinguishable elements on regular lattice possess strong redundancy. The total number of possible pairs of antennas, corresponding with the total number of measured visibilities, scales with the square of the number of elements in the array. On the other hand the number of unique baselines (given by the number of unique separation vectors between all possible pairs of antennas) scales only linearly with the number of elements. Since all baselines made of identical elements and spanning the same distance vector should measure the same signal, we can use this to back out calibration factors of individual antennas, without ever knowing anything about the actual sky signal. The solution is unique up to intrinsic degeneracies of the system corresponding to the overall shift, scaling in amplitude, and translation of the sky signal (i.e. applying a phase gradient across the u-v plane). This calibration procedure is known as redundant calibration and has been worked in detail in \cite{2010MNRAS.408.1029L} and \cite{2016ApJ...826..181D}.

Unfortunately, it soon became clear that real arrays are much less redundant for the simplest form of redundant calibration to be sufficient. For example \cite{2003.08399} studied redundancy in HERA and found that real-life non-redundancy produces spurious temporal structure in gain solutions. \pa{ Other works such as \cite{2101.02684, Joseph_2018_BIAS,wieringa1992investigation,Zheng_2014,Zheng_2016,Orosz_2019} discuss the shortcomings and possible solutions of redundant calibration in both real telescope arrays and simulations.}

In some sense, these findings indicate that redundant calibration is the victim of its own success. A possible way to look at the problem is that it is not that the arrays are too non-redundant, but that they are too sensitive. They are sensitive enough that the particularities of individual elements produce big enough effects that a good fit cannot be found for the data. One possible approach has been studied in \cite{2017arXiv170101860S}: a pair of nearly, but not perfectly, redundant baselines will have very strong, but not perfect, correlation between measured visibilities. These slight decorrelations can be propagated self-consistently using a quadratic estimator formalism to allow stable solutions and almost by construction, a good fit. This process has the advantage of gracefully dealing with outlier antennas: there is no need to cut them out if we can instead model them as such. A more recent paper \cite{2004.08463}
proposed a unified approach to sky-based and redundant calibration, in which both can be used concurrently in a self-consistent Bayesian model, assuming we have a model of the telescope and its calibration uncertainties. 

\pa{In} this paper we take a different approach: instead of describing every single antenna with one complex number, a single gain, we are looking for a description in terms of multiple numbers per element that can be adjusted in order to achieve a good fit as well as give some physical insight into the type of imperfections. Note that this does not break the basic premise of redundant calibration: the number of unknowns still scales linearly with the number of antennas, while the number of measurements scales quadratically. So, for arbitrarily complex description of per-antenna non-idealities, the system will be over-constraint for large enough array (as long as the number of free parameters per antenna is finite).

\pa{We will focus on two of the canonical errors} in a redundant radio array\pa{, pointing errors and geometric errors. Some of the other canonical errors include beam shape and beam size.} The pointing errors refer to the fact that the beams of individual element are not aligned (i.e the telescope points away from the zenith in a transit array configuration) and geometric errors, which means that the entire dish is displaced from its nominal lattice positions affecting the effective baseline lengths. These are the kinds of errors that we study, but we do not explicitly fit for them. Instead, we build a general pixelized description of the response of each element that is arbitrarily fine-resolved given by a tunable parameter $M$. \pa{Ideally, our scheme would also be able to deal with beam shape and beam size issues, however we focus on geometric and pointing errors.}

\section{Problem set-up}

Let us consider a general interferometric array observing, for simplicity, in a single polarization mode.  We consider a pair of antennas pointing at the zenith and observing in a narrow frequency range.  The noiseless observed visibility for pairs of antennas $i$ and $j$ in the flat-sky approximation is given by 
\newcommand{\vu}{\mathbf{u}}
\newcommand{\vx}{\mathbf{x}}
\newcommand{\FT}[1]{\mathcal{FT}[#1]}
\newcommand{\oV}{V^o}
\newcommand{\tB}{\tilde{B}}
\begin{multline}
  V_{ij} = \int  I(\theta) e^{-2\pi i (\vx_i - \vx_j)  \cdot \theta } B_i(\theta) B_j^*(\theta)  d^2\theta =\\
  \FT{I B_i B_j^*}(\vx_i - \vx_j)
\end{multline}
where $\vx_i$ is the geometric position of antenna $i$ measured in wavelengths, $I(\theta)$ is the intensity signal coming from the sky as a function of angle on the sky, $\theta$ and $B_i$ is the beam of antenna $i$. The operator $\FT{X}$ denotes the Fourier transform of $X$.  The signal $I(\theta)$ is real and in principle spans the entire sky. The beams $B_i$ are in general complex and compact in Fourier space (i.e. on the u-v plane). They correspond to the complex amplitude response to an incident uniform plane wave just above the aperture. 

Multiplication in real space corresponds to convolution in Fourier space. Addition of the random noise component $\epsilon$ gives the general equation for the observed visibility for a pair of antennas
\begin{equation}
 \oV_{ij} = \int U(\vx_i-\vx_j+\vu) (\tB_i \circledast \tB_j^\dagger)(-\vu) d^2\vu + \epsilon
\label{eq:main}
\end{equation}
Where the uv-plane $U = \FT{I}$ contains the image of the sky (in Fourier space), and $\tB_i = \FT{B_i}$ are beams representation in this domain. Variable $\epsilon$ is a random variable describing the noise realization. We have used notation $\tB^\dagger(\vu) = \FT{B^*}(\vu) = \FT{B}(-\vu)$, i.e. complex conjugation of a real space quantity results in mirroring across the origin in the Fourier domain.  
The crucial insight is that beams are compact in Fourier space, essentially corresponding to the physical extent of the dishes and thus $\tB_i \circledast \tB_j^\dagger$ is also compact. Therefore, the limits of integration in Equation \ref{eq:main} need to extend only as far as support of $\tB_i \circledast \tB_j^\dagger$. 

It is instructive to compare this equation with the one that is typically used to set-up redundant calibration:
\begin{equation}
    \oV_{ij} = U_{i-j} g_i g_j^* + \epsilon,
\label{eq:redundant}
\end{equation}
where $i-j$ indexes redundant baselines corresponding to the baseline vector $\vx_i - \vx_j$.
If we set our beams to be the same, scaled by only a complex gain factor for each dish, namely $B_i (\theta) = g_i B (\theta)$,
we find that the Equation \ref{eq:main} reduces to the Equation \ref{eq:redundant} with 
\begin{equation}
    U_{i-j} = \int U(\vx_i-\vx_j+\vu) (\tB \circledast \tB^\dagger)(-\vu) d^2\vu
\end{equation}
In other words, the redundant calibration assumes that all dishes have exactly the same sky response that are only allowed to vary by an overall complex gain, which is assumed to be coming from electronics, amplifiers, etc. It reconstructs these gains, but also the uv-plane information, \emph{after the beam convolution}. 

But in general, massive redundancy in compact arrays allows us to go beyond this approximation (we will quantify this in the next section). The main idea is to start with Equation \ref{eq:main} as the main ingredient and reconstruct both $\tB$ for each antenna and the uv-plane $U$. It is important to note, from the beginning, that we cannot unpick the variations in complex gain coming from electronics from those arising from the imperfect dish. Therefore, action of a gain $g_i$ is equivalent to multiplying the entire beam by the same factor.

\subsection{Discretization}

We will try to build an intuition that will prepare us for formulation of general redundant calibration in steps. In short, we will try to build a redundant calibration that will replace the parameters $U_{i-j}$ and $g_i$ for the standard redundant calibration in the Equation \ref{eq:redundant} with a new, larger, but still finite number of parameters that can still be solved for by using the inherent redundancy in the system but can allow for some freedom in non-redundancy.

In order to proceed, we need to express the integral of the Equation (\ref{eq:main}) as a sum over a finite number of degrees of freedom.  Our approach is to pixelize the beam maps $\tB_i$ and choose the pixelization in the uv-plane $U$ that matches the beam pixelization in a way that expresses Equation (\ref{eq:main}) as a sum that is cubic in input parameters without entailing any interpolation.

For concreteness, let us consider an interferometric array on regular square grid of size $N_s \times N_s$ with a total of $N_a=N_s^2$ antennas. We will measure all distances in the units of wavelength and take the lattice spacing to be $L$. We start by introducing the ``oversampling'' parameter $M$. The beam map is pixelized in $(2M+1)^2$ pixels, ie. a central pixels and $M$ additional pixels on each side. For $M=0$ the beam is reduced to a single pixel and the method reduces to a redundant calibration. The requirement on the odd number of pixels in the beam description is there to ensure that a single pixel description of each beam is a natural limit of the problem.

\begin{figure}[h]
  \centering
  \includegraphics[width=\linewidth]{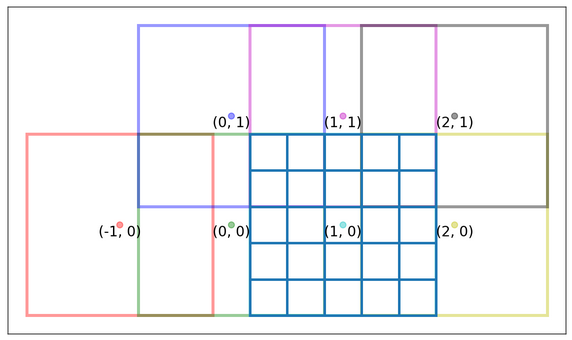}
  \caption{Illustration of the pixelization scheme for $M=1$. Labeled points illustrate the lattice points on the $u-v$ plane for a square close-packed array. For example $(1,0)$ corresponds to baselines one lattice spacing apart in the E-W direction and $(0,1)$ is the same for N-S baslines. $(0,0)$ is the origin of the $u-v$ plane and corresponds to single-dish observations. Individual beam is described by $3\times 3$ pixelized grid, so $(\tB \circledast \tB^\dagger)$ is the $5 \times 5$ pixelized grid shown in blue around $(1,0)$. Making prediction for a particular $(1,0)$ baseline amounts to multiplying the values of the blue beam convolved map on the blue grid and the corresponding points in the $u-v$ plane and summing them up. See text for further discussion.   }
  \label{fig:pixelization}
\end{figure}
The convolution of two beams $(\tB_i \circledast \tB_j^\dagger)$ is of the size $(4M+1) \times (4M+1)$, which sets the natural pixel size to be $L/(2M+1)$ for both the beam grid and the $U$ grid. We can therefore re-write the Equation \ref{eq:main} in a pixelized form schematically as
\pa{
\begin{equation}
    V_{ij} = \sum_{m=-2M}^{+2M} \sum_{n=-2M}^{2M} U_{i_o+m;j_o+n} (\tB_i \circledast \tB_j^\dagger)_{mn},
\end{equation}}
where $i_o$ and $j_o$ are the grid offsets corresponding to the $i-j$ redundant baseline \pa{(the baseline probed by beam $\tB_i$ and $\tB_j$)} on the uv-plane, \pa{and $m$, $n$ are the pixel indices over the convolved beam.} Note that the sum is over $(4M+1)\times (4M+1)$ corresponding to the convolution of two $(2M+1)\times(2M+1)$ sized beam maps. This is illustrated in the Figure \ref{fig:pixelization}. We urge reader to spend some time trying to understand this Figure as this pixelization is essential for the understanding of our approach. Consider the shortest E-W baseline of a closely-packed dish array. Dishes have diameter $D$ so the distances between pieces of reflector surface on the same dish vary between $0$ and $D$. The possible distances between pairs of pieces of reflector surface on two dishes vary between $0$ (at points where the dishes nominally touch) and $2D$. In the formalism, this is encoded by the beam being $(2M+1)^2$ pixels in size for an individual beam and $(4M+1)^2$ for the convolution of the two beams. This then defines the extent on the $u-v$ plane to which this particular baseline is sensitive to. In the limit of large $M$ it comes arbitrarily close, but never reaches the origin of the $u-v$ plane (corresponding to the monopole signal) and similarly gets arbitrarily close but never reaches two lattice spacings (corresponding to sensitivity doe to pieces of reflector surface that are furthest apart). In Figure \ref{fig:pixelization} this is illustrated with the blue grid.  \as{This is the correct and expected behavior for the flat sky approximation. We discuss the non-flat sky case further below.}

In Figure \ref{fig:pixelization} we also show squares around other nominal lattice pointings. We see that most $u-v$ points are being probed by multiple baselines, except those on precise grid spacings. This severely limits the degeneracies present in the problem \pa{which we will further comment on in a later section}. We also note that we need to describe just one half of the u-v plane with other half given by the reality of the observed field. Together with rules about Fourier transforms of conjugated fields, this ensures that prediction for the antenna pair $i,j$ is always a complex conjugate of the prediction for the antenna pair $j,i$. 

\subsection{Phased-up array interpretation }

To continue building intuition about the pixelization, consider the pixel $m$ in the antenna $i$ and pixel $n$ in the antenna $j$. The visibility for the pairs of dishes $i$ and $j$ is given by 
\begin{equation}
  V_{ij} = \sum_{\substack{m \in \mbox{\tiny \ Beam\ }i \\ n \in \mbox{\tiny \   Beam\ }j}} U_{\Delta (i,m)-(j,n)} \tB_{i;m} \tB^*_{j;n},
\end{equation}
where the index to $U$ schematically implies the $u-v$ plane distance between pixel $m$ on antenna $i$ and pixel $n$ on antenna $j$.   This equation now has the form of Equation \ref{eq:redundant} (with gains replaced by pixels inside the beam response),  but added over all the pixel pairs formed by the two beams. In other words, the pixelization is in effect modelling the redundant array made of $N_s\times N_s$ antennas as a grid of $N_{\rm eff} = (2M+1)N_s \times (2M+1)N_s$ independent antenna elements which are phased up in blocks of $(2M+1) \times (2M+1)$.

\as{We can now return to the question of the applicability of this technique to the flat sky approximation. The redundant calibration is typically derived without considering going beyond the flat sky approximation, but it is clear that it applies generally. An identical pair of antennas will see nominally the same signal and so the values on the $u-v$ plane can be thought as simply keeping track of the true visibility corresponding to a certain baseline orientation without specifying how these are related to the actual sky emissions. The same argument applies to our case, but now involves an approximation. Our model implies that the individual beam can be composed from a phased up elements of equivalent antennas. Consider two sub-baselines of one such baseline pair of the same length, one stretching from central to central pixel and one stretching from two side pixels. While the model can account for different gains, it cannot account for different ``primary fields of view'' of two such finite size sub-elements, especially far away from the pointing centre.  Therefore, this description will be necessarily approximate. At the same time, it is also true that in the limit of arbitrary fine pixelization, where the sub-elements are smaller than the wavelength, the ``primary fields of view'' of imagined sub-elements encompass the entire sky and can thus describe a general beam profile. However, the number of unknowns in that case will explode beyond what can be reasonably fit.
}

\subsection{Redundancy factor}

The total number of unknowns to be determined is thus given by i) the total number of pixels required to describe all the beams $N_a (2M+1)^2$ and ii) the pixels required to describe uv-plane, which are given by the number of redundant baselines in the effective phased-up array $2N_{\rm eff}(N_{\rm eff}-1) = 2 N_s (2M + 1) (2 M N_s + N_s - 1)$.  The total number of measurements, on the other hand goes as the total number of baselines with measured visibilities, which is given by $N_a(N_a-1)/2$. Thus, we can construct the redundancy factor \pa{for a square close-packed array}
\begin{equation}
    r = \frac{\sqrt{N_a}\left(N_a - 1\right)}{2\left[\left(2M +1
    \right)\left(6M\sqrt{N_a} + 3\sqrt{N_a} - 2\right)
  \right]}
\label{eq:red}
\end{equation}
which describes the ratio of the number of measured quantities divided by the number of unknowns. The $M=0$ case reduced to the standard redundant calibration scheme. Note that since we are assuming a square array, the $\sqrt{N_a}$ is always an integer. When $r\geq 1$, the system is over-constrained and we can hope to constrain it, subject to known degeneracies. Also, note that the ratio in Equation (\ref{eq:red}) is for the particular case of square lattice. For an imperfectly filled array or a hexagonal close-packing, the ratio would be different. But note that this formalism does not even require antennas to be on the lattice. It can be calculated for any array and even an irregular but sufficiently packed array will have $r>1$.

\begin{figure}
    \centering
    \includegraphics[width=.9\linewidth]{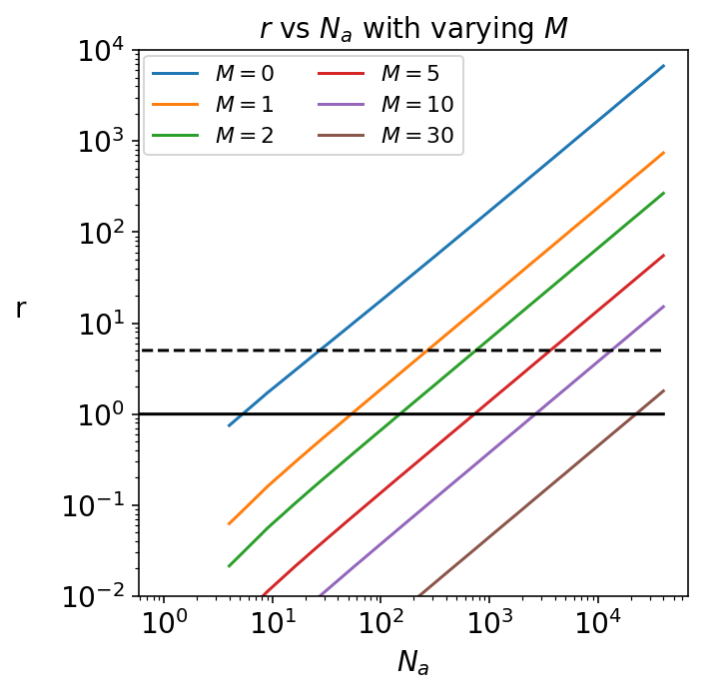}
    \caption{Redundancy ratio as a function of $N_a$ with varying M. Solid black line is $r=1$ and dashed black line is $r=5$ corresponding to significant redundancy. }
    \label{fig:ratiovsNa}
\end{figure}
In Figure~\ref{fig:ratiovsNa} we plot $r$ as a function of $N_a$ for a couple of values of $M$. We see that the for the traditional redundant calibration, even very small arrays containing tens of antennas are sufficiently redundant. HIRAX, with a $32\times32$ array can theoretically model to up to $M=5$ but realistically more likely up to $M=2$. In what follows we will focus on the $M=1$ case which is perhaps sufficient for the current arrays such as HERA. In this exploratory work, we have not done an explicit calculation for a cylinder array configuration like that of CHIME, although that would be a straightforward generalization.

\subsection{Comparison with other approaches}
\as{We are not the first paper to extend the minimal redundant calibration. In fact, in the seminal paper \cite{2010MNRAS.408.1029L} they considered an option to Taylor expand the $u-v$ plane to first order around the lattice positions. In other words, the model for visibilites is expanded from Equation \ref{eq:redundant} to
\begin{equation}
  \oV_{ij} = \left(U_{i-j} + \nabla U_{i-j}\cdot \Delta \mathbf{x}_{i,j}\right) g_i g_j^* + \epsilon,
\end{equation}
where $\Delta \mathbf{x}_{i,j}=\mathbf{x}_i-\mathbf{x}_j$ is the vector difference between the nominal and actual baseline length, so that the term in brackets corrects for the slightly non-redundant baseline length. One can then self-consistently fit for all the derivatives $\Delta U_{i-j}$ and individual antenna positions.}

\as{Given the discussion above (see Equation \ref{eq:redundant}), the standard redundant calibration assumes the primary beams to be perfectly aligned and so the quantity $U_{i-j}$ is the actual $u-v$ plane after the convolution with primary beams. Therefore the Taylor expansion method can deal with geometrical errors (see below) exactly (within the limits of Taylor expansion), but not with pointing errors.
We see that an almost natural extension of this approach is to extend the Taylor expansion in the beam mis-pointing direction as follows:}
\begin{equation}
  \oV_{ij} = \left(U_{i-j} + \nabla U_{i-j}\cdot \Delta \mathbf{b}_{i,j} + \Phi_{i-j} \cdot \Delta \mathbf{p}_{i,j}\right) g_i g_j^* + \epsilon,
\end{equation}
\as{where $\Phi$ is a 2D field defined on the grid points that represents a response of the system to mis-pointing and $\Delta \mathbf{p}_{i,j} = (\mathbf{p}_i+\mathbf{p}_j)/2$ is the mean mis-pointing\footnote{While seemingly counter-intuitive, the mispointing goes with the mean displacement, not the difference. If one dish points off-centre to the north, while the other off-centre to the south, then to first order they will see the same signal, which is suppressed by factor that is quadratic in mispointing. If they both mispoint north, the signal will change in proportion to the mean mis-pointing.}. The system will still be over-constrained for a sufficiently large array. }  \as{We have not pursued this approach in the name of generality. Since sufficiently big arrays can be very redundant, the idea is to let the system be capable of absorbing arbitrary imperfections, even those that fall in the standard expectation of geometry, pointing and beam size errors. }

\as{An alternative approach has also been described in \cite{2017arXiv170101860S}. The idea here is to deal with imperfect redundancy by sweeping it under the rug, statistically. The main idea is that in a perfectly redundant array, the redundant baselines would measure the same number; as we dial up the array imperfections, the nominally redundant arrays will not measure the same values, but only highly correlated ones. One can self-consistently write a statistical quadratic model that tracks those correlations and is able to gracefully describe the array non-redundancy. The main issue with this approach is that the differences are not exploited for their information content, they are simply treated as random fields rather than information that can be used to understand in precisely what way the array is imperfect.}

\section{Numerical implementation}

In previous sections we have presented a general method for modeling the signals from an imperfect interferometer. The methods contains the ``resolution'' parameter $M$ which controls how fine are the pixels which describe the response of each individual beam. $M=0$ corresponds to standard redundant calibration and $M\rightarrow \infty$ is a completely general description. The hope is that a relatively small $M$ will be sufficient to describe typical level of non-redundancy. The purpose of this section is to what extent this is true and our logic is to attempt to use $M=1$ on relatively small arrays to focus on pointing and geometry errors. The path we follow is to assume a concrete perfectly redundant array observing a given sky and then see how the general redundant calibration performs compared to a standard redundant calibration as we introduce beam non-redundancies.

\subsection{Model Beam and its Imperfections}

Our model beam is a circularly symmetric tapered filled circle in the Fourier space:
\begin{equation}
   \tilde{B}_c(r) =  1 - \frac{1}{1 + e^{-2\cdot\frac{r - r_0}{\Delta r}}}.
\end{equation}
\as{This is a more realistic than the usual approximations which usually assume a purely Gaussian or purely Airy disc beams and suffices for this exploratory work.}  Note that this is description of the Fourier transform of the beam, rather than the beam itself, i.e the  real space beam becomes an Airy disc in the limit of $\Delta r$ going to zero.  The reason why we specify the beam in Fourier space is partly because this is the input quantity we need for making predictions, but also more importantly it corresponds to space with a compact beam representation. \as{Since antenna is physically compact, the complex electric field in a plane just above the antenna is compact. This approximation breaks down in two ways in realistic instruments. First, when we Fourier transform this to obtain the shape of the beam on the sky we are implicitly making a flat-sky approximation. Second, coupling to neighbouring elements will make the antenna physical size bigger than what we mechanically consider to be the antenna element. }

\begin{figure}
    \centering
    \includegraphics[width=.8\linewidth]{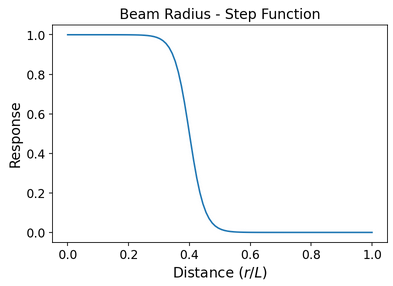}
    \caption{Radius response function with $R=0.4 L$ and taper$=0.05 L$. }
    \label{fig:beam}
\end{figure}

We use units of lattice spacing, so radius $r_0=0.5 L$ would correspond to dishes that just touch.  In our simulation we use $r_0=0.4 L$ and $\Delta r = 0.05 L$. We show the beam in real and Fourier space in Fig \ref{fig:beam}. \as{The beam shows the expected sidelobes in real space.}

\begin{figure}
    \centering
    \includegraphics[width=\linewidth]{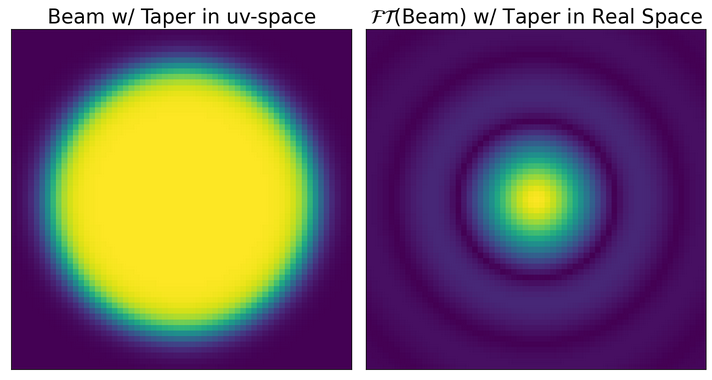}
    \caption{ \pas{2D visualization of the beam} $\tilde{B}$ used on the left and its Fourier transform, the actual primary beam $B$, on the right. The left image spans the $uv$-plane while the right is in the x-y plane. In the limit of no taper, the beam on the right would be an Airy disc pattern. }
    \label{fig:2dbeam}
\end{figure}

In this work we focus on to most common types of errors encountered in redundant arrays: the pointing errors (the center of the beam of the dish is misplaced) and geometry errors (the dish is not at its nominal location on the lattice). Pointing errors are displacement of the beam in the Fourier space and thus correspond to applying a phase gradient across the complex beam response of a single dish. The geometry errors on the other hand and just simple displacement of the dish in the Fourier domain. A full expression for the beam is thus given by
\begin{equation}
\tilde{B}(x,y) = A \tilde{B}_c\left(\sqrt{(x-x_o)^2+(y-y_o)^2} \right) e^{-i (p_x x + p_y y)},
\end{equation}
where $A$ corresponds to the overall complex gain factor, $(x_o,y_o)$ is the geometric error and $(p_x,p_y)$ is the pointing error. We draw both the overall complex gains from a Gaussian distribution centered around $1+0i$ with a variance of $\sigma_g$. Geometric and pointing errors are drawn from 2D Gaussians with 1D variances of $\sigma_g$ and $\sigma_p$ respectively.

\begin{figure*}
\centering
\includegraphics[width=.7\textwidth]{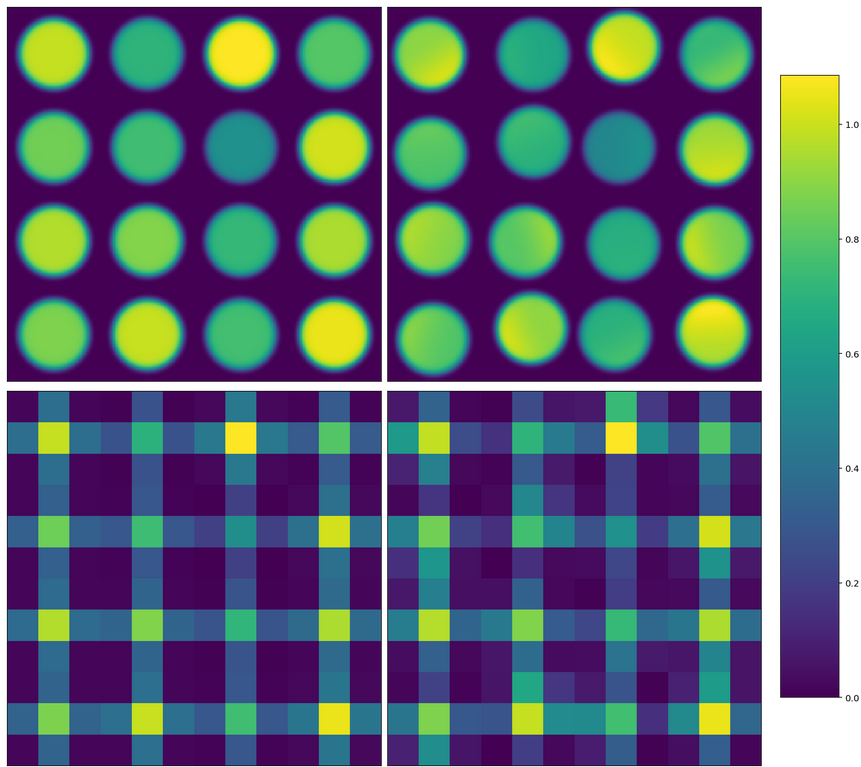}
\caption{Example image of a $4\times 4$ array in $uv$-space with a perfectly redundant system on the left and a perturbed version of the same systems on the right. The top images are generated using $M=30$ and bottom are the same system but downscaled to $M=1$ resolution. The complex gain of each pixel in the telescope dish is turned into an RGB value for this image with the color bar referring to the overall amplitude ranging from 0 to slightly greater than 1.
  The perturbations for the right-side graphs were generated with geometric errors at $\sigma_g=.1L$ and pointing errors at $\sigma_p=.5$. Pointing errors which lead to phase changes are modeled as a gradient on each beam and geometry errors are modeled as circle offsets.
}
    \label{fig:two_arrays}
\end{figure*}

In Figure \ref{fig:two_arrays} we illustrate a perfect and an imperfect array with the types of errors discussed above.

\subsection{Simulating Signal}

\as{We simulate the signal relying on a flat sky approximation.  We do so using two approaches.} In the first approach, we simply generate data using a large value of $M$. We use $M=14$ (corresponding to beam images of 29$\times$29 pixels). In this case we take the true values of the $u-v$ plane as random gaussian variates, corresponding to a white-noise signal on the sky. We have checked that increasing $M$ beyond the chosen value used does not affect our results.

In this approach, the code using for generating the signal and fitting is essentially the same with only the value of $M$ being different. On one hand this is elegant, but it is also prone to potential bug inadvertently canceling out between data generation and fitting. Moreover, the realistic skies are often dominated by a few bright sources.

\begin{figure}
    \centering
    \includegraphics[width=\linewidth]{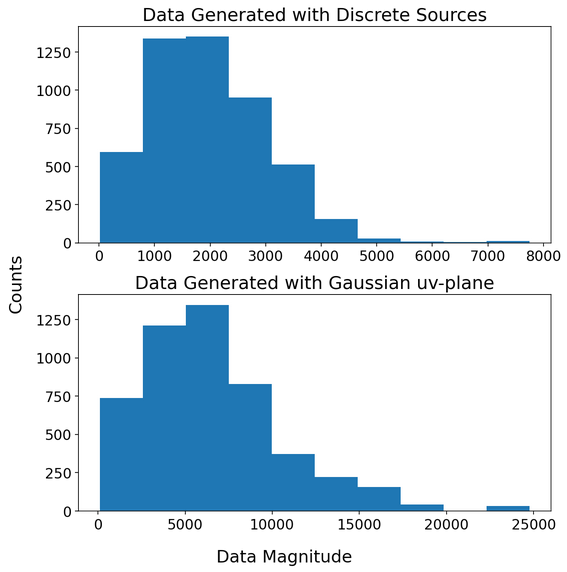}
    \caption{\pa{Histogram comparing the magnitude of observed visibilities generated between the two methods. There are 100 discrete sources used to model the sky. The overall statistics of the two are comparable alleviating concern about the sky simulation with $M$. }}
    \label{fig:data_histograms}
\end{figure}

In the second approach we generate the signal as a sum over discrete sources. This is to take into account the fact that some sources are considerably brighter than the others. For each source and baseline we:
\begin{itemize}
\item Calculate the primary beam response at the location of the source for both dishes. Given pointing errors this is different for each baseline.
\item Directly calculate the response of the interferometer for the baseline length spanned by a given pair of dishes (taking into account geometry errors).
\end{itemize}
We use sources whose fluxes are randomly drawn uniformly in log from 1 to 1000. 
\pa{A comparison between the two methods can be found in Fig~\ref{fig:data_histograms}
. Aside from an overall scale factor the two are quite similar.}

In both approaches, once the noiseless visibilities are calculated, we multiply visibility by the total complex gain contribution $g_i g_j^*$, where $g_i$ is the gain for dish $i$.  Note that the method is completely general with respect to individual gains, i.e. they can be perfectly absorbed into beam description. Therefore we keep the variance small to avoid dealing with solver falling into a local minima.

Finally, we add complex  noise $\epsilon$ drawn from a Gaussian with variance $\sigma_\epsilon^2$. Since the only quantity that matters is the level of noise compared to the level of signal, we will report our results in terms of signal-to-noise ratio (SNR).  We define the SNR per visiblity as
\begin{equation}
  {\rm SNR}^{2} =  \frac{1}{N_{\rm visibilities}} \sum_{\mbox{\tiny baselines }i-j} \left(\frac{|V_{i-j}^2|}{\sigma_\epsilon^2}\right)^{2}  
\label{eq:snr}
\end{equation}

There are thus 6 additional degrees of freedom per each dish: 2 for pointing error, 2  for geometry error and 2 for overall gain. Even with $M=1$, each beam is described by 9 complex numbers (i.e. 18 degrees of freedom) so there is reasonable hope that the description is sufficient.
However, on purpose, we decided to consider the types of errors that are not perfectly reproducible by our model to asses its flexibility. 
Had we instead decided to model dish imperfections as \pa{random deviations the size of a pixel on our $(2M+1)^2$ grid, say as dirt 1/9th the size of our dish for $M=1$}, our model would be guaranteed to perform better \pa{ compared to redundant calibration.} 

\section{Solving for $\tilde{B}_i$ and $U$}
\label{sec:solving}
In this work we use simple iterative solvers for beams $\tilde{B}_i$ and the true $u-v$ plane $U$. Since we are focusing on the modelling side, i.e. how well the solutions perform, these solvers are not designed to be either particularly stable or fast. 
\pa{That being said, some notes and improvements are listed in the relevant sections below.}

In general, we are trying to maximize the log likelihood of the model,  which is equivalent to minimizing the $\chi^2$, given by
\begin{equation}
  \chi^2 = \sum_{\mbox{all pairs }i,j} \frac{(V_{ij}^o - V_{ij}^p(\tilde{B}, U))^2}{\sigma_\epsilon^2}, \label{eq:chi2}
\end{equation}
where $V_{ij}^p$ denotes the predicted visibilities which are a function of all beam parameters and $u-v$ plane values. At each step we optimize for either visibilities or beam parameters.

\subsection{Visibility}
To solve for visibilities, $U$, we rely on the fact that observed visibilities are linear in the input visibilities.
\begin{equation}
  V_{i}^o = M_{ik}U_k + \epsilon \label{eq:visib_est}
\end{equation}
\pa{Here, $i$ iterates over observed visibilities and $k$ over the input, or true, visibilities. While the specific values will depend on the geometry of the array, $i \sim N_a^2$ and $k \sim N_a^2M^2$.}
The matrix $M$ is quite sparse, but because beams overlap, it is not a block matrix. In other words,  neighbouring baselines do probe some of the same sky signals as illustrated in the Figure \ref{fig:pixelization}.
Matrix $M$ depends on all beams and so we assume the current best guess for the beams (which improve with every iteration). We rely on \pa{Scipy\cite{2020SciPy-NMeth} for actual calculations. Specifically,} to calculate $(\tB_i \circledast \tB_j^\dagger)$.  we use \verb|scipy.signal.convolve| and to solve the sparse system we rely on \verb|scipy.sparse.linalg.lsqr|. The rest is rather painful but otherwise straightforward housekeeping. For interested reader we point at some of the more subtle technical issues in Appendix \ref{app:tech}.

\pa{At this point we can use any number of methods to solve a linear equation with \texttt{scipy.sparse.linalg.lsqr} being the most direct in application. We can also employ any number of tools to better solve our linear system such as the Wiener filter which we motivate in the Regularization section. Using Equation \ref{eq:minimal_variances}}
we can use \verb|scipy.sparse.linalg.spsolve| to solve the system, avoid any tough inversions, and although there is more overhead, it is ultimately faster than directly solving Equation (\ref{eq:visib_est}). Notably, employing a Wiener filter is also considerably better at reducing $\chi^2$ per iteration over \pa{\texttt{ scipy.sparse.linalg.lsqr}} which makes the filter a worthwhile implementation.

\subsection{Beams}
While Equation \ref{eq:main} is nominally quadratic in the beams, this is not an issue in our actual problem, because we do not consider auto-correlation signal. To solve for the beam $B_i$, we fix all the remaining beams and the solved $u-v$ plane $U$, so that
\begin{equation}
V_{k}^o = M_{kl} \tilde{B}_{i,l} + \epsilon
\label{eq:beam_lin}
\end{equation}
Here index $k$ runs over all visibilities that depend on the beam $i$ (i.e all baselines that contain antenna $i$) and $l$ over all pixels of beam $\tilde{B}_i$. This is a dense system that we solve using \verb|scipy.optimize.lsq_linear| separately for each beam $i$. \pa{While this can be embarrassingly parallel, we found improvements to $\chi^2$ per iteration when done sequentially. After solving for $\tB_i$, the updated solution is used when writing down $M_{kl}$ for $\tB_{i+1}$ and so on. Of course this can be circumvented with some clever distributed computing which we leave for the future.}

\section{Perfect degeneracies}
\label{sec:reg}
The standard redundant calibration has perfect degeneracies spanning 4 degrees of freedom:

\begin{itemize}
\item Multiply gains by complex factor $\alpha$ and divide sky signal by $\alpha \alpha^*$ (this is often split into the overall amplitude degeneracy and the overall phase degeneracy);

\item Translate the sky by a vector $\mathbf{p}$ and apply a consumerate phase gradient across the gain solution;
\end{itemize}

The same degeneracies continue to exist in our case. One would naively expect that we also have a similar set of per-element degeneracies, however, these are not present, in general, because neighbouring antennas actually measure the many of the same points in the $u-v$ plane, thus introduce ``interlocking'' of the $u-v$ plane solutions.

However, we have a different kind of degeneracy present. We know that if an array is truly redundant, then one needs only  $\sim 3N_s^2$ numbers to describe the data. So if the array is actually fully redundant, we are free to pick any ``shape'' of the beam $\tilde{B}$ for $M>0$ and still have sufficient freedom in the $U$ array to form a model that gives precisely the same predictions.  In the other limit, if the array is really non-redundant then this degeneracy disappears. Therefore this is not really a model degeneracy, but a degeneracy associated with a perfectly redundant array solutions which are, from a mathematical perspective,  pathological. This is analogous to solving a matrix equation ${\rm M}x = y$, which for a general matrix M is solvable by $x = {\rm M}^{-1} y$, unless matrix M is singular. 
\pa{Removing these degeneracies using polarization data, as suggested by \cite{Dillon_2018} should be plausible but tenuous with non-redundant arrays. With our array formalism it is possible to tackle this issue but left as a future endeavour.}

In practice, the presence of noise will instead use the extra model freedom to ``fit the noise'' and find a nominally better solution as we will describe in the Results section. 
A formally correct way would be to perform a strict Bayesian model comparison, where we weight the solutions by the Bayesian evidence in favor of a certain model: if the model with $M>0$ can fit the data equally well as the standard redundant calibration, then the standard redundant calibration would be strong favored due to having many fewer priors.

\section{Regularization}
\label{sec:regularization}

To prevent overfitting, we implement \pa{2 regularization schemes} as follows. We first introduce a prior on the beam parameters that attempts to pull the beam solution back to the fiducial, redundant beams, \pa{and also include an option to minimize the variance on solved visibilities.}
The total likelihood \pa{with the beam prior} is thus given by
\begin{multline}
  \log L = -\frac{1}{2}\chi^2 + \\
  \sum_{{\rm beam\ } i} \sum_{{\rm pixel\ }k}\left(-2 \log \sigma_B -\frac{|(\tilde{B}_{ik} - \tilde{B}_{k}^0)|^2}{2\sigma_B^2} \right),
\end{multline}
where $\chi^2$ is given by Equation \ref{eq:chi2} and $\tilde{B}^0$ corresponds to \pas{the beam prior. As stated above we choose this to be the perfect, unperturbed beam.}

The $\sigma_B$ describes typical deviation from perfect beams. Taking derivative of the $\log L$ with respect to to $\sigma_B$ (with other parameters fixed), one gets
\begin{equation}
  \sigma_B = \sqrt{\frac{1}{2N} \sum_{{\rm beam\ } i} \sum_{{\rm pixel\ }k}  \left(|(\tilde{B}_{ik} - \tilde{B}_{k}^0)|^2\right)}\label{eq:sigmaB}
\end{equation} with $N$ being the total number of pixels over all beams, $(2M+1)^2 N_a$.

If the array is close to truly redundant, the system can achieve a good fit by floating beams towards nominal beam values and lowering $\sigma_B$ (and thus achieve a large likelihood improvement through the normalization term $-\log \sigma_B$). However, for a non-redundant array, it is better to raise $\sigma_B$ and instead improve the $\chi^2$.
\pa{This prior will bias the beams towards having a magnitude of 1, or whatever the normalization of the perfect beam is set to. To counteract this bias, we introduce an overall gain for each dish which applies to each pixel. Our solver implements the \textit{omnical}\cite{2003.08399} method to iteratively solve for overall gains.}

Finally, we put a prior on visibilities. In a more complete treatment, we would solve for the visibilities power spectrum at the same time as we solve for the visibilities map, similar to the Gibbs' sampling approach to CMB (see \cite{2004PhRvD..70h3511W}). In our case we just add a penalty term to the solved visibilities in the form of \begin{equation}
    f = \lVert\Sigma_b^{-1}\cdot \left(M_{ik}U_k - V_i^o\right) + \Sigma_v^{-1}\cdot U_k\rVert
\end{equation} with respect to $U_k$ where $\Sigma_b^{-2} = N^{-1}$ is the noise covariance and $\Sigma_v^{-2} = S^{-1}$ is the data covariance. In our simulations we assume both are diagonal to simplify calculations. This leads to a Wiener filter. Following the form written in \cite{Tegmark_1997} we specifically solve for visibilities using the following linear equation 
\begin{equation}
    \left[S^{-1} +(M_{ik})^\dagger N^{-1}M_{ik}\right]U_k = 
    \left((M_{ik})^\dagger N^{-1}\right)V_i^o.
    \label{eq:minimal_variances}
\end{equation}

\section{Recap of the iterative solver}

To recap, we maximize the likelihood with iterative steps. In order, each step involves fitting for: 
\begin{enumerate}
\item The visibilities by solving either Equation \ref{eq:visib_est} or \ref{eq:minimal_variances}.
\item The beam shape by solving \ref{eq:beam_lin} and incorporating the prior parameter, $\sigma_B$, by adding rows to the bottom of our linear system.
\item The overall per-antenna gain parameters $A_i$ via \textit{omnical}\cite{2003.08399}.
\item Finally, the beam parameter $\sigma_B$ which amounts to calculating the variance of beam solutions with respect to their nominal values as written in \ref{eq:sigmaB}.
\end{enumerate}

\begin{figure*}
    \centering
    \includegraphics[width=.95\textwidth]{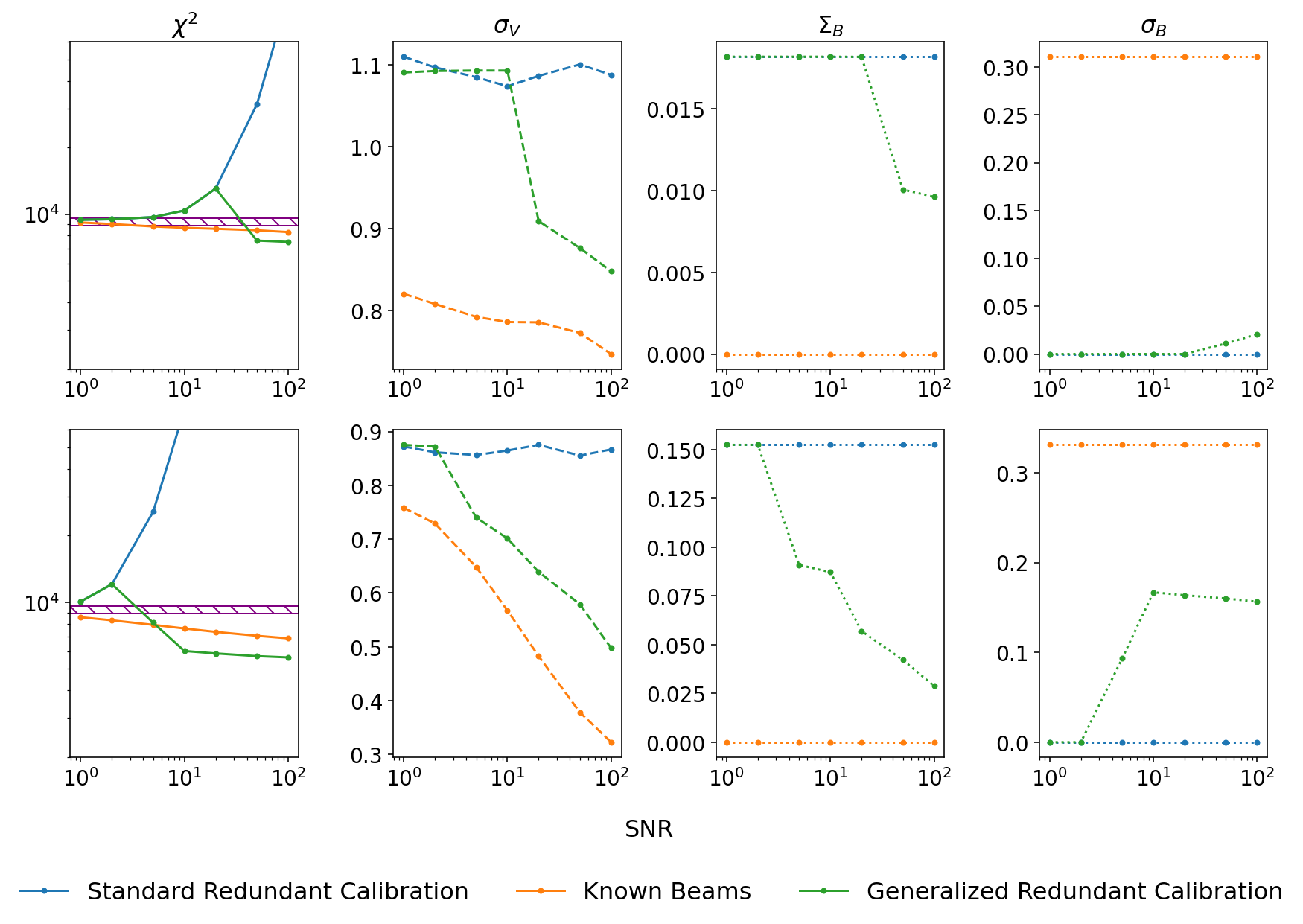}
    \caption{Comparison of our algorithm (green) against perfect knowledge of beam shape (orange) and also against an assumption of a perfectly redundant array (blue).
      Blue and orange have fixed beams, at the prior and truth respectively, and can only vary the visibilities and overall gain factors. As such these two methods are forms of standard redundant calibration.
    Our algorithm solves using M=1 and the data is generated with M=1, meaning that the model is a perfect description of the data.  The top row corresponds to a weak departure from perfect redundancy with the geometric and pointing errors set to $.01$, while the bottom row is for relatively strong departure from redundancy with geometric and pointing errors set to $0.1$.
    The plots in the first column show $\chi^2$ for each model to see if we can recreate the data. The purple dashed box is
    the 5 percentile to 95 percentile bound of the expected $\chi^2$ for generalized redundant calibration with no regularization. $\chi^2$ below this box is indicative of overfitting.
    The second column plots the variance between the solved and true visibilities, $\sigma_V$, to see if we are accurately recovering the true sky -- since we are generating and solving using M=1 we
    can compare the solved to true values directly. The third column plots the variance between the solved and true beam shapes, labelled $\Sigma_B$, to see if we recover the correct array. As the blue and orange lines use a fixed beam shape (the beam prior and true beams respectively), their values for $\Sigma_B$ are constant.  
    The fourth column plots the variance between the solved and prior beam shape, $\sigma_B$ from the Equation~(\ref{eq:sigmaB}).
    }
    \label{fig:M3_2fig}
\end{figure*}

\begin{figure*}
    \centering
    \includegraphics[width=.95\textwidth]{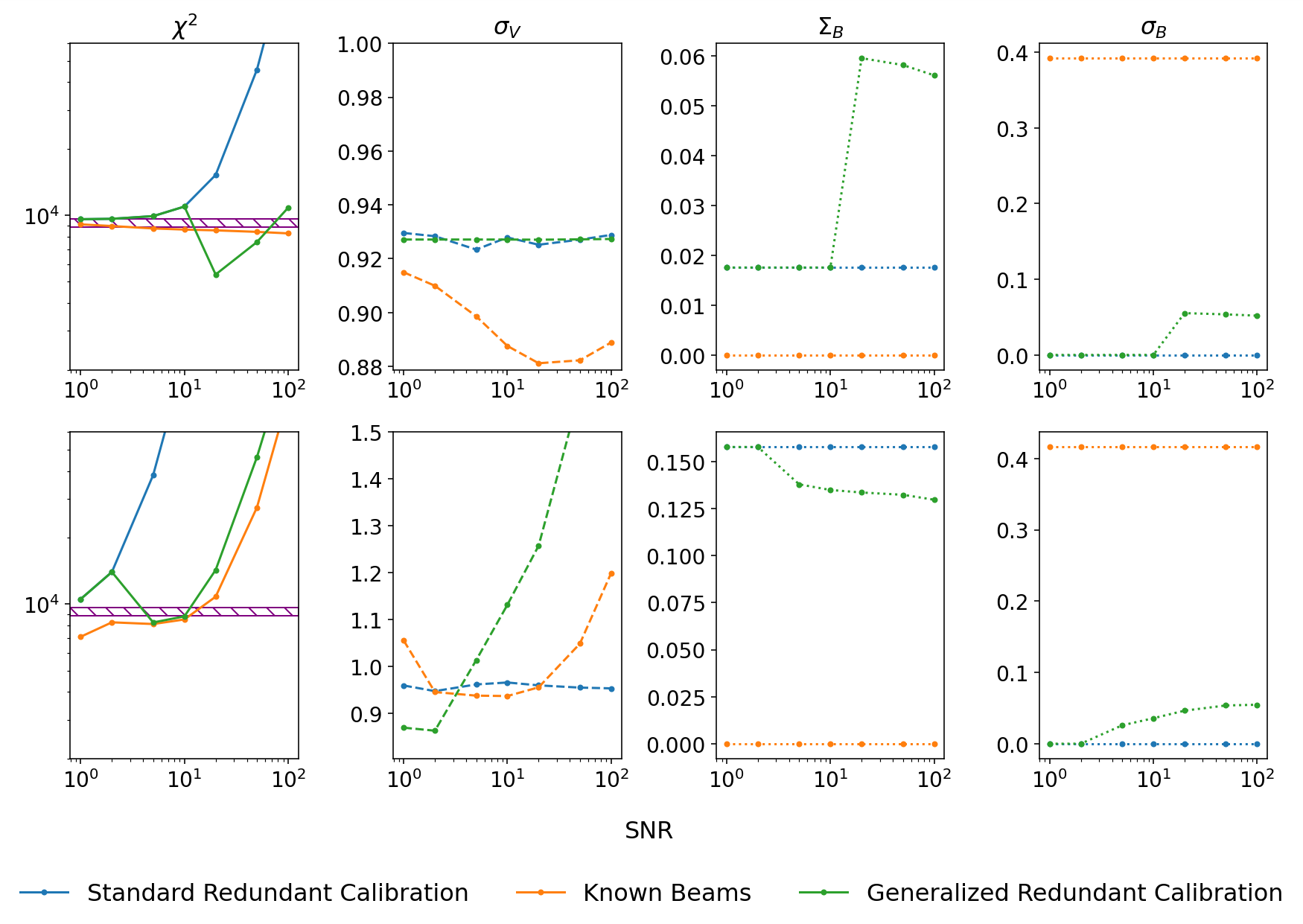}
    \caption{Same as Figure~\ref{fig:M29_2fig} but with data generated using $M=14$ which is essentially indistiguishable from the continuous model. The data were still fitted using $M=1$ model, which is now an approximate model. For $\sigma_v$, the central pixels from the fitted $uv$ data have been compared with the averaged $uv$-plane pixels from the corresponding area from the model representing the truth. See text and caption to Figure~\ref{fig:M29_2fig} for complete description.}
    \label{fig:M29_2fig}
\end{figure*}

\section{Results}





The purpose of telescopes is to image the sky. If we have an imperfect description of the physical effects that affect the instrument's response and calibration vectors that are only approximately correct and whose values are ``effective'', it should not matter as long as the map of the sky is faithfully reproduced. 
Therefore the method that will best recover the input $uv$-plane is the most successful method. 

We have condensed our results into two Figures \ref{fig:M3_2fig} and \ref{fig:M29_2fig} that succinctly summarize our results regarding the method.  For both Figures, the upper-row of plots corresponds to rather modest geometry and pointing errors, while the bottom row contains considerably worse errors. All figures show various quantities plotted as a function of per-visibility SNR (see Equation \ref{eq:snr}). The left column shows the values of $\chi^2$: high values of $\chi^2$ indicate a poor fit, while lower values a better fit; values below the shaded region indicating the expected $\chi^2$ given degrees of freedom indicate over-fitting. The central left column shows the goodness of the $uv$-plane recovery as measured by the variance between the true and recovered $uv$-plane, labelled $\sigma_V$. This is only done for the $uv$-plane pixels that lie on the nominal beam array lattice positions where the SNR is concentrated. The lower the value, the more faithful recovery of the $uv$-plane a certain method is giving. The central right column plots show the variance of the recovered beams relative to the true beams which we call $\Sigma_B$, while the variance between recovered beams and prior beams, $\sigma_B$, is plotted in the right-most column. The lower the $\Sigma_B$ the closer we are in recovering the true values of the beam. The value of $\sigma_B$ on the other hand tells us not only how far the beams are from their nominal (unperturbed) values but also how far our method thinks they are in average, since $\sigma_B$ is at the same time a regularization parameters (see Section \ref{sec:reg}).  Remember that the overall phase factor is set in a separate parameter and that the central beam pixels was set to unity -- central pixels are not used in the variance calculation. 

Three different colors correspond to three different methods. In blue we plot the standard redundant calibration which we implement by simply forcing beams to be at their nominal, un-perturbed values and recover just the gains and the $uv$-values. Note that it is still not exactly the same as standard redundant calibration, because the recovery is on the $uv$-plane quantities before convolution with the beam. In orange we plot the same, but this time we instead fix the beams to their true value by assuming that somehow these were independently calibrated with infinite precision. This sets the upper limit of how well a generalized redundant calibration should perform. Finally in green we show the results of our method in its full self-calibrating glory.

We start with a simplified case, Figure \ref{fig:M3_2fig}, in which we generate the data using exactly the same theory as we fit (the data is generated with $M=1$ and fit with $M=1$).
In this case, the fitted theory is by construction a perfect description of the data. The purpose of this exercise is to isolate the effects of overfitting from effects of using an approximate theory.

At low signal-to-noise, the method cannot really detect with certainty that the beams are different from their unperturbed values. The regularization thus follows the beam prior, resulting in a small $\sigma_B$ and a relatively large $\Sigma_B$.

As the signal-to-noise increases, the tension becomes significant, and the method relaxes $\sigma_B$, allowing beams to unstick from their unperturbed prior values. $\Sigma_B$ therefore undergoes a transition to smaller values with beam values moving towards their true values.  In the bottom panel, as SNR increases, our algorithm asymptotes towards  $\Sigma_B = 0$, matching the orange line. The process works better for the more non-redundant array, because  while non-redundancy is a source of noise generically, in our case it is also a source of signal.  In $\chi^2$ this is manifested as $\chi^2$ increasing with signal-to-noise and then ``snapping'' down towards lower values when $\sigma_B$ is relaxed. We find that even in this case the statistical system is prone to over-fitting. This indicates that further regularization is likely warranted in the $uv$-plane sector.  At the same time, we see the redundant calibration simply unable to explain the data with the $\chi^2$ monotonically increasing away from good fit with either increasing SNR or the non-redundancy of the array.

In the second column we see consistent behaviour. The orange line with known beams performs best and its variance on the visibilities monotonically decreases as the SNR increases -- the fidelity of the solution increases. The standard redundant calibration is systematically limited and as the $\chi^2$ increases it is simply unable to improve its solution.
Overall, the generalized redundant calibration is between the blue and orange lines with issues of falling into local minima at low SNR.
But for sufficiently high SNR it out-performs the standard redundant calibration as expected. In never performs quite as well as the case with known beams, because the latter utilizies significant extra amounts of information.

In Figure \ref{fig:M29_2fig} we show the same set of plots for a more realistic cases where we simulated data with $M=14$ and recover them with $M=1$ theory as before.
Here the beam are described in a true theory by a 29x29 matrix which we average to a 3x3 matrix in cells in order to get ``true'' beams.
Similarly we average the true $uv$-plane into a reduced resolution $uv$-plane that we compare with recovered values.
The details of how this is performed matter, but largely do not affect our results.
The right hand side plots showing the $\sigma_B$ behaviour are largely unchanged.
However, we see that the $\chi^2$ plot keeps increasing as the SNR increases.
This indicates that we are fitting a model that is actually unable to fit the data.
The main assumption of this paper, namely that a low number of extra degrees of freedom in individual beams parameterized as pixels will be sufficiently flexible to describe a general geometrical and pointing errors has proven to be insufficient at SNR per measurement of over $\sim$10.
As we increase the signal-to-noise, the recovery of the map (second panel) first improves, but then starts to get worse.
Interestingly we find that for large level of array non-redundancy all methods perform worse at higher SNR; this is possible when methods fall into the wrong local minima or where the imperfections of the model are such that better formal fits actually performs worse in the quantities of interest. But it is also possible that our recipe to convert the $M=14$ truth into $M=1$ pixelized $uv$-plane are just too simplistic. In this case, the implication is that we have a solution that is formally good but with a poorly understood relation to the underlying truth.

\section{Conclusions}
We have presented a new method for calibrating imperfect redundant arrays. The method is a derivative of redundant calibration and models each independent beam element as a phased-up array of $(2M+1) \times (2M+1)$ sub-elements, each with its own complex gain factors. In the limit of large $M$, the method is capable of modeling any array, by having a complete freedom to represent the response of each dish as pixelized $(2M+1) \times (2M+1)$ complex beam response. In the limit of $M=0$, the method reduces to the standard redundant calibration.

To avoid fitting for the noise we have attempted a regularization scheme that models the departures from the perfect beams using a Guassian with diagonal scatter. The magnitude of scatter $\sigma_B$ is a fitted parameter. As expected, we found that when signal-to-noise is low, and the data is sufficiently well described by the standard redundant calibration, the beams solutions relax to their priors and $\sigma_B$ becomes essentially zero. In this limit, the system has less tendency to fit for the noise, although we find that the $\chi^2$ remains too low and and noise fitting remains an issue. When the signal-to-noise is sufficient to detect non-redundancy, the value of $\sigma_B$ rises and for sufficiently non-redundant array, the solutions approach those without regularization. 

In this paper we have focused on methodological aspects of this method, namely, is the method capable of producing good fits to the data. In practice, while this might be true, the very high dimensionality of this problem makes finding of these solutions difficult. We found that, unless we start with a good approximate guess, the method is likely to fall into a local minima.  Therefore, in order to make this method practically usable, it is necessary to first find efficient minimizers. Moreover, the method currently works with a single sky snap-shot and should be generalized to time-stream data. 

Unfortunately, we have found that low $M$ configurations are not good at describing generic array errors. The high $M$ configurations will likely perform considerably better, however, given the increased model complexity in that case is even more likely to suffer from overfitting and preponderance of local minima. The correct solution to these issues is to employ a much more sophisticated marginalization scheme than the maximum likelihood scheme employed in this work. 

While we have focused on nearly redundant-arrays, our method is in trivially generalizable to only partially redundant arrays in arbitrary configuration. A fixed value of $M$ defines a grid with $D/(\lambda M)$ in the $u-v$ plane. Any array containing dishes (even heterogeneous ones!), whose response with respect to some arbitrary original can be satisfactorily described on the $u-v$ plane in this grid can be in general fit with generalized redundant calibration. Of course, this model is interesting only if the array has a sufficient redundancy that the number of unknowns does not exceed the number of observed visibilities, since otherwise it is capable of trivially explaining any measurement.

We found that the main downside of this method is that it is non-trivial to connect the measured values to the underlying quantities of interest. We have seen that the central $uv$-plane pixels (i.e. those on the nominal lattice positions) are well recovered, but this has not been investigated in any detail.  In comparison, in the standard redundant calibration, the interpretation of the fitted $uv$-values is clear: they are exactly the values of the true $uv$-plane convolved with the appropriate beam responses. Unfortunately, in generalized redundant calibration however, the actual beams are not made up from identical sub-beams. Therefore, it is non-trivial to precisely connect the recovered $(2M+1)\times(2M+1)$ pixelized beam back to actual pointing and geometric offsets. 

The beauty of the proposed scheme is that is it very general, especially for large array that could afford to go beyond $M=1$. On the other hand, if we did know that the dominant errors are pointing and geometry errors, one could design a model that would fit explicitly for those. In that case, the contents of the u-v plane could be modeled by the values and its spatial derivatives at the lattice points. We leave extended comparisons of these methods for the future.

\section*{Acknowledgements}

Authors acknowledge useful feedback obtained during the PUMA
collaboration seminar series.  This work was supported in part by the
U.S. Department of Energy, Office of Science, Office of Workforce
Development for Teachers and Scientists (WDTS) under the Science
Undergraduate Laboratory Internships Program (SULI).

\appendix

\section{Technical considerations in fitting}
\label{app:tech}
Since we are probing nearby baselines, it is possible that we probe the conjugate of a known visibility, i.e. while looking at the baseline $(0, 1)$ we will pick up some amount of signal from the baseline $(-\frac{1}{3}, 1)$.
It is important that we are careful to enforce that 
\begin{equation}
    U_{(-\frac{1}{3}, 1)} = \left(U_{(\frac{1}{3}, -1)}\right)^*.
\end{equation}
This means that even if we construct a large sparse matrix $M$ (from Equation~\ref{eq:visib_est}) with each row corresponding to the probed baseline $\alpha = (i,j)$ with coefficients from $P_{ij}$ (the convolution of two beams), we cannot simply solve for $\vec{U}$ using this system.

Thus the correct way to implement Equation~\ref{eq:visib_est} would be to split our $\vec{U}$ into real and complex components and doing the same with our data.
Explicitly, given a generic $a, b \in \mathbb{C}$, with $a = a_1 + a_2i$ and similarly for $b$, we can say 
\begin{align}
    a\cdot b  &= (a_1b_1 - a_2b_2) + (a_1b_2 + a_2b_1)i, \\
    a^*\cdot b  &= (a_1b_1 + a_2b_2) + (a_1b_2 - a_2b_1)i.
\end{align}
So it simply comes down to flipping the sign on a few of the coefficients of our convolution, $P_{ij}$ for some of the baselines.

While the housekeeping for visibility solving boils down to changing signs and splitting our complex data into two parts, writing out Equation~\ref{eq:beam_lin} to solve for the beams is a bit more involved.

To write our 2D convolution as a matrix product we can use a blocked Toeplitz matrix.
For each $\tB_i$ we can write it's effect in convolution as a matrix $\beta_i$ by writing it as a blocked Toeplitz matrix, and for $\tB_j^\dagger$ we simply have $\beta_j^\dagger$.
Then our data can be written as 
\begin{align}
    V_{ij} &= \left(\tB_j^\dagger \circledast  \tB_i \right) \cdot \vec{U}_\alpha \\
    &= \vec{U}_\alpha \cdot \left(\beta_j^\dagger \tB_i\right)\\
    &= \left(\vec{U}^T_\alpha \beta_j^\dagger \right) \cdot \tB_i \\
    &= M_{ij}\tB_i
\end{align}
Here we have a fixed $i$ and iterate over the other beams $j$.
Unlike the visibility solver, there is no relation between any of the beams and thus we are able to solve each one as an independent linear system.

\bibliography{references}

\end{document}